\documentclass[twocolumn,showpacs,amsmath, amssymb, prl, natbib]{revtex4-2}

\usepackage[utf8]{inputenc}
\usepackage{graphicx}
\usepackage{amssymb}
\usepackage{amsmath}
\usepackage{natbib}
\usepackage[capitalise]{cleveref}
\usepackage{color}

\newcommand{\be}{\begin{equation}}
\newcommand{\ee}{\end{equation}}
\newcommand       \bea          {\begin{eqnarray}}
\newcommand       \eea          {\end{eqnarray}}
\newcommand       \apjl         {ApJL}

\newcommand       \mnras        {MNRAS}

\newcommand       \araa      {ARA\&A}

\begin{document}

\def\red#1{\textcolor{red}{\textbf{#1}}}
\def\blue#1{\textcolor{blue}{\textbf{#1}}}

\title{Enhanced Extreme Mass Ratio Inspiral Rates into Intermediate Mass Black Holes}
\author{Ismail Qunbar}\email{ismail.qunbar@mail.huji.ac.il}
\author{Nicholas C. Stone}\email{nicholas.stone@mail.huji.ac.il}
\affiliation{Racah Institute of Physics, The Hebrew University, 91904, Jerusalem, Israel}


\begin{abstract}
Extreme mass ratio inspirals (EMRIs) occur when stellar-mass compact objects begin a gravitational wave (GW) driven inspiral into massive black holes.  EMRI waveforms can precisely map the surrounding spacetime, making them a key target for future space-based GW interferometers such as {\it LISA}, but their event rates and parameters are massively uncertain.  One of the largest uncertainties is the ratio of true EMRIs (which spend at least thousands of orbits in the {\it LISA} band) and direct plunges, which are in-band for at most a handful of orbits and are not detectable in practice.  In this paper, we show that the traditional dichotomy between EMRIs and plunges -- EMRIs originate from small semimajor axes, plunges from large -- does not hold for intermediate-mass black holes with masses $M_\bullet \lesssim 10^5 M_\odot$.  In this low-mass regime, a plunge always has an $\mathcal{O}(1)$ probability of failing and transitioning into a novel ``cliffhanger'' EMRI.  Cliffhanger EMRIs are more easily produced for larger stellar-mass compact objects, and are less likely for smaller ones.  This new EMRI production channel can dominate volumetric EMRI rates $\dot{n}_{\rm EMRI}$ if intermediate-mass black holes are common in dwarf galactic nuclei, potentially increasing $\dot{n}_{\rm EMRI}$ by an order of magnitude. 
\end{abstract}

\date{\today}
\maketitle

\noindent {\it Introduction}-- The advent of gravitational wave (GW) astronomy has opened a new window into the universe of compact objects (COs).  Present-day ground-based interferometers are sensitive to GWs produced by mergers of stellar-mass COs \cite{Abbott+16, Abbott+17, Abbott+21}, but over the next decades, space-based interferometers such as {\it LISA} will access new portions of the GW spectrum \citep{AmaroSeoane+22}.  A primary science target for {\it LISA} is the extreme mass ratio inspiral (EMRI), a GW signal produced when a stellar-mass CO slowly inspirals into a massive black hole (MBH).  The slow, quasi-adiabatic evolution of an EMRI can build up enormous signal-to-noise ratio \cite{HilsBender95, AmaroSeoane18}, precisely mapping the spacetime of the central MBH \cite{Ryan95}, and testing general relativity by measuring its multipole moments \cite{Ryan97}.

At present, the astrophysical origins and rates of EMRIs are highly uncertain.  The classical (``loss cone'') channel for EMRI production relies on two-body relaxation of orbits in galactic nuclei, which feeds COs to the central MBH on high-eccentricity orbits \citep{HilsBender95, SigurdssonRees97, HopmanAlexander05, BarOrAlexander16}.  Alternate EMRI production mechanisms include low-eccentricity inspirals of tightly bound COs deposited by the Hills mechanism \cite{Miller+05}, orbital migration of COs through active galactic nucleus accretion disks \citep{PanYang21, DerdzinskiMayer22}, and scatterings off MBH binaries \citep{Mazzolari+22}.  Of these channels, the loss cone is the best-studied, but even here, volumetric EMRI rates are uncertain by at least three orders of magnitude \citep{Babak+17}, depending sensitively on the unclear MBH occupation fraction in dwarf galaxies \citep{Miller+15}, rates of CO depletion and production in galactic nuclei \citep{AharonPerets16}, the relaxation of nuclear cusps \citep{AmaroSeoanePreto11}, and the (possibly top-heavy) stellar mass function in nuclear environments \citep{Bartko+09, Lu+13}.

A crucial complication to the loss cone channel for EMRI production is the dichotomy between true EMRIs, which spend thousands of orbits in the {\it LISA} band, and plunges, which are far harder to detect because they spend at most a handful of orbits in-band.  The distinction between these two fates is traditionally understood to arise from a competition between two-body relaxation, which causes a random walk in CO orbits, and GW dissipation, which powers a deterministic inspiral \citep{HopmanAlexander05}.  Past theoretical loss cone calculations find that the large majority of COs consumed by MBHs spend their final years on orbits where two-body relaxation dominates GW emission, and thus terminate in a plunge.  The ratio of plunges to EMRIs has been estimated to be $\mathcal{R} \sim 10-100$ \citep{Merritt15, BarOrAlexander16}, at least in the high-density nuclei that likely dominate volumetric capture rates \citep{Merritt15}.

In this {\it Letter}, we show that the classic understanding of the plunge/EMRI dichotomy is incomplete, and that the loss cones of intermediate-mass black holes (IMBHs) cannot achieve $\mathcal{R} \gg 1$.  In contrast, the traditional dichotomy remains correct for supermassive black holes (SMBHs), at least if they are not spinning.  We first demonstrate this point with approximate analytic arguments, and then we quantify it more precisely using Monte Carlo (MC) orbit simulations.  Finally, we explore and summarize the broader implications of this result for EMRI rates.  Throughout the paper, $G$ is Newton's constant and $c$ the speed of light.

\noindent {\it Inspirals and Plunges}-- We consider as an idealized model a nuclear star cluster residing in the potential of an MBH with mass $M_\bullet$.  The mass density of surrounding stars is assumed to be spherically symmetric, $\rho(r)=\rho_{\rm inf}(r/r_{\rm inf})^{-\gamma}$, where $r_{\rm inf}$ is the influence radius (defined as the point where the enclosed stellar mass equals $M_\bullet$) and $\rho_{\rm inf}$ is the mass density there.  The gravitational potential is assumed to be Keplerian, $\Phi = -GM_\bullet /r$, over the radii of interest, and thus the local velocity dispersion is $\sigma^2(r) = GM_\bullet (1+\gamma)^{-1} / r$.  The GW inspiral time (at leading post-Newtonian order) for a CO of mass $m$ on a highly eccentric orbit of semimajor axis $a$, eccentricity $e$, and pericenter $q=a(1-e)$ is \cite{Peters64} 
\begin{equation}
t_{\rm GW} = \frac{3 \times 2^{7/2}}{85} \frac{a^{1/2}q^{7/2} c^5}{G^3 M_\bullet^2 m},
\end{equation}
while the angular momentum relaxation time for such an orbit (i.e. the characteristic timescale for the random walk produced by uncorrelated two-body scatterings to change angular momentum by an order unity amount) is
\begin{equation}
t_{\rm AM} = \frac{2k \sigma^3(r)}{G^2 n(r) \langle m^2 \rangle \ln \Lambda} (1-e).
\end{equation}
Here $k \approx 0.34$ \cite{BinneyTremaine08}, the stellar number density $n(r) = \rho(r) / \langle m \rangle$, $\ln \Lambda \approx \ln(0.4 M_\bullet/\langle m \rangle)$ is the Coulomb logarithm, and $\langle m \rangle$ and $\langle m^2 \rangle$ are the first and second moments of the stellar mass distribution (respectively). 

There is a critical value of specific angular momentum that determines the minimum parabolic pericenter (i.e. innermost bound circular orbit) for orbits around a non-spinning MBH, $J_{\rm IBCO} = 4GM_\bullet / c$.  This can also be understood as a minimum pericenter \footnote{In Schwarzschild coordinates, $q_{\rm IBCO}$ is located at $r=4 R_{\rm g}$, but the gauge-dependent nature of distances in general relativity makes a direct mapping to Newtonian coordinates problematic.  In this paper we make the less arbitrary but still imperfect choice of first mapping integrals of motion ($J$) into a Newtonian framework, and then computing $q_{\rm IBCO}$}, $q_{\rm IBCO} = 8 R_{\rm g}$, where $R_{\rm g} = GM_\bullet / c^2$ is the gravitational radius.  When a CO's orbit has $J \gg J_{\rm IBCO}$, $t_{\rm GW} \gg t_{\rm AM}$.  As $J\to J_{\rm IBCO}$, both timescales will drop, but $t_{\rm GW}$ will drop faster.  There is thus a critical semimajor axis that separates true EMRIs (where GW dissipation takes over orbital evolution) from plunges \cite{HopmanAlexander05}, which we derive by setting $t_{\rm GW}=f t_{\rm AM}$:
\begin{equation}
a_{\rm c}^{3-\gamma} = \frac{85}{3\times 2^{5/2}} \frac{kf}{(1+\gamma)^{3/2}\ln\Lambda} \frac{G^{5/2}M_\bullet^{7/2}m}{\langle m^2 \rangle c^5 q^{5/2} n_{\rm inf} r_{\rm inf}^\gamma} \label{eq:aC}. \\
\end{equation}
For example, if $\gamma=7/4$ (a Bahcall-Wolf cusp \cite{BahcallWolf76}), then $a_{\rm c}\approx 5.8\times 10^{-3} r_{\rm inf} (x_{\rm IBCO}/8)^{-2} (\mathcal{M}/10)^{4/5} (f \kappa/7.4)^{-4/5}$, where we defined a dimensionless pericenter $x \equiv q/R_{\rm g}$ (evaluated at its minimum value $x_{\rm IBCO}$), a mass combination $\mathcal{M} \equiv m \langle m \rangle / \langle m^2 \rangle$, and a collection of constants $\kappa \equiv (1+\gamma)^{3/2}(3-\gamma) \ln \Lambda$ that is normalized to a fiducial value $\kappa \approx 74$ (as for $M_\bullet = 10^6 M_\odot$, $\gamma=7/4$).  We also set $f=0.1$ throughout the paper to achieve better agreement with numerical simulations (see e.g. Fig. \ref{fig:SMass} below).

In past literature, this critical semimajor axis has been either estimated or assumed to represent an absolute and fairly rapid transition, with all COs entering the MBH event horizon with $a \lesssim a_{\rm c}$ EMRIs, and all with $a \gtrsim a_{\rm c}$ plunges.  However, this dichotomy is not universal, and in particular is not very firm when considering sufficiently small MBHs.  Past work seems to have neglected the possibility that an orbit which begins its random walk with $a \gg a_{\rm c}$ can end it with $a \lesssim a_{\rm c}$ due to strong GW dissipation once $J \sim J_{\rm IBCO}$.  We quantify whether this possibility can occur by asking if a CO can lose enough energy in a single orbit to push its semimajor axis $a < a_{\rm c}$.  The energy dissipated in a single orbit can be expressed, at leading post-Newtonian order, as 
\begin{equation}
\Delta E_{\rm GW} = \frac{85\pi}{12\sqrt{2}} x^{-7/2}mc^2 \frac{m}{M_\bullet}.
\end{equation}
Equating $a_{\rm new} = GM m / (2\Delta E_{\rm GW})$ to $a_{\rm c}$, we find that a ``failed plunge'' can occur when 
\begin{equation}
M_\bullet \lesssim M_{\rm c} = \left(\frac{85\pi  mc^2 a_{\rm c}}{6\sqrt{2}G x_{\rm IBCO}^{7/2}} \right)^{1/2} \label{eq:Mc} 
\end{equation}
$M_{\rm c} \approx 10^{4.9}M_\odot (x_{\rm IBCO}/8)^{-11/4} (r_{\rm inf}/{\rm pc})^{1/2} (m/10 M_\odot)^{4/5}$ if $\gamma=7/4$, $\kappa \approx 74$ and $\langle m^2 \rangle / \langle m \rangle = M_\odot$.  Since $\ln \Lambda$ depends weakly on $M_\bullet$, fixing $\kappa$ slightly overestimates $M_{\rm c}$.  We can likewise write
\begin{equation}
    x_{\rm c}= \left(\frac{85 \pi mc^2 a_{\rm c}}{6 \sqrt{2} G M_\bullet^2} \right)^{2/7} ,
\end{equation}
the minimum dimensionless pericenter for a failed plunge to occur.  Note that $x_{\rm c}$ is a monotonically decreasing function of $M_\bullet$. For MBHs with $M_\bullet \gtrsim M_{\rm c}$, the standard picture is basically correct, and $a_{\rm c}$ represents a transition between two clearly distinct asymptotic regimes.  But when $M_\bullet \lesssim M_{\rm c}$, the situation is quite different for orbits with $a \gg a_{\rm c}$, as these may transition from plunge to EMRI trajectories.  Since orbital evolution prior to the onset of an EMRI is stochastic, these failed plunges are not guaranteed to happen, they are merely possible.  If the final step in the CO's random walk takes it from $x_{\rm n-1}  > x_{\rm c}$ to $x_{\rm n} < x_{\rm IBCO}$, then a plunge will still occur.  It is only if $x_{\rm IBCO} < x_{\rm n} \lesssim x_{\rm c}$ that the naive plunge fails and can become an EMRI, an outcome we refer to as a ``cliffhanger EMRI'' hereafter.  

Following past literature \citep{HopmanAlexander05}, we define the EMRI fraction $S(a)$ to be the ratio of EMRIs to all captures (EMRIs and plunges).  $S(a) \to 1$ when $a /a_{\rm c} \to 0$.  If $M \gtrsim M_{\rm c}$, we expect to recover the classical picture, where $S(a) \to 0$ as $a/a_{\rm c} \to \infty$.  But if $M_\bullet \lesssim M_{\rm c}$, then we expect $S(a) \to S_\infty > 0$ as $a/a_{\rm c} \to \infty$.

\noindent {\it Monte Carlo Simulations}-- To test the approximate arguments made above, we perform orbit-averaged Monte Carlo (MC) orbit simulations that account for (i) GW dissipation and (ii) stochastic perturbations from other stars in the nuclear star cluster.  In our MC simulations, the per-orbit energy loss to GW emission is \cite{Peters64}
\begin{equation}
\Delta E_{\rm GW} = -\frac{64 \pi G^{7/2} M_\bullet^{3/2} m^2 (M_\bullet + m)}{5 c^5 a^{7/2} (1 - e^2)^{7/2}} \left(1 +\frac{73}{24} e^2 + \frac{37}{96} e^4\right)
\end{equation}
and the corresponding per-orbit angular momentum loss is 
\begin{equation}
\Delta L_{\rm GW} = -\frac{64 \pi G^{5/2} M_\bullet^{3/2} m^2 (M_\bullet + m)^{1/2}}{5 c^5 a^2 (1 - e^2)^2} \left(1 + \frac{7}{8} e^2\right)
\end{equation}
These deterministic losses of energy and angular momentum combine with the stochastic effects of weak stellar scatterings to determine the overall evolution of the test compact objects we are considering. We treat weak (small-angle) scatterings in the usual diffusive approximation \citep{Merritt13}, with diffusion coefficients presented in Supplementary Material for completeness.  Throughout this paper we neglect stellar diffusion due to resonant relaxation, as past studies have shown this to be relatively unimportant for both action-space fluxes near $x \sim x_{\rm IBCO}$ and for overall EMRI rates \citep{BarOrAlexander16}.

We implement the MC simulations with the algorithm of Ref. \cite{ShapiroMarchant78}, which uses a variable timestep for computational efficiency and self-consistently handles cross-correlated variability in $E$ and $J$.  At each timestep, we increment energy and angular momentum by amounts
\begin{align}
    \delta E =& \Delta E_{\rm GW} N + N\langle \Delta E \rangle_{\rm o} + N^{1/2}y_1 \langle (\Delta E)^2 \rangle_{\rm o}  \\
    \delta L =& \Delta L_{\rm GW} N  + N\langle \Delta L \rangle_{\rm o} + N^{1/2}y_2 \langle (\Delta L)^2 \rangle_{\rm o}.
\end{align}
Here each simulation timestep is equivalent to $N$ orbits with a period $P(E)$.  We adjust $N$ dynamically to ensure that $|\delta E/E| \ll 1$ and $|\delta L/L| \ll 1$, subject to the constraint that $N\ge 1$.  The dimensionless random variables $y_1$ and $y_2$ are generated so that they have means $\langle y \rangle=0$, dispersions $\langle y^2\rangle=1$, and a cross-correlation $\langle y_1 y_2 \rangle = \langle \Delta E \Delta L\rangle_{\rm o}^2 / (\langle (\Delta E)^2 \rangle_{\rm o} \langle (\Delta L)^2 \rangle_{\rm o})$.  Further details are contained in Supplementary Material.  In all cases, we take $\langle{m}\rangle=M_\odot$, $\langle{m^2}\rangle=M_\odot^2$, and $r_{\rm inf}=11 ~{\rm pc}~(M_\bullet / (10^8 M_\odot))^{0.58}$, as is appropriate for low-mass cusp galaxies \citep{StoneMetzger16}.  Large increases in $\langle m^2 \rangle$ could occur for a cusp rich in stellar mass black holes, and would decrease $M_{\rm c}$ by a factor of a few.

Our MC simulations identify a transition between two different regimes of EMRI behavior, as can be seen in Fig. \ref{fig:SMass}.  For $M_\bullet \gtrsim 10^5 M_\odot$, we are in the ``classical'' regime, where COs that begin their random walk with an initial semimajor axis $a_0 \ll a_{\rm c}$ terminate their random walk as EMRIs, and those that begin with $a_0 \gg a_{\rm c}$ terminate their random walk with direct plunges.  
However, for $M_\bullet \lesssim 10^5 M_\odot$, we find a novel regime, at least for $a \gg a_{\rm c}$ initial conditions.  At these lower MBH masses, the asymptotic behavior as $a_0 / a_{\rm c} \to \infty$ is quite different, with $S(a_0) \to S(\infty) > 0$. This numerical result is in excellent agreement with Eq. \ref{eq:Mc}, which gives $M_{\rm c}\approx 10^{4.9}M_\odot$, in alignment with Fig. \ref{fig:SMass}.

In the MC calculations used to produce Fig. \ref{fig:SMass}, we considered the case where all COs are stellar-mass black holes with $m = 10 M_\odot$.  In Fig. \ref{fig:SCO}, we fix $M_\bullet = 10^4 M_\odot$ and instead vary the CO mass.  We see that for $m=1.4 M_\odot$ neutron stars, the behavior of $S(a)$ returns to the classical picture, while for $m = 50 M_\odot$ black holes, the MC results deviate even further from past work, with $S(\infty)$ increasing further. This behavior can be understood in terms of the competition between GW emission and angular momentum relaxation that sets the asymptotic, $a_0 \to \infty$ behavior.  As $m\to \infty$, rates of AM relaxation remain largely unchanged (so long as $m \ll M_\bullet$ remains true) but GW emission at fixed pericenter increases steadily, favoring the large single-passage energy losses needed to initiate a cliffhanger EMRI.

\begin{figure}
\includegraphics[width=80mm]{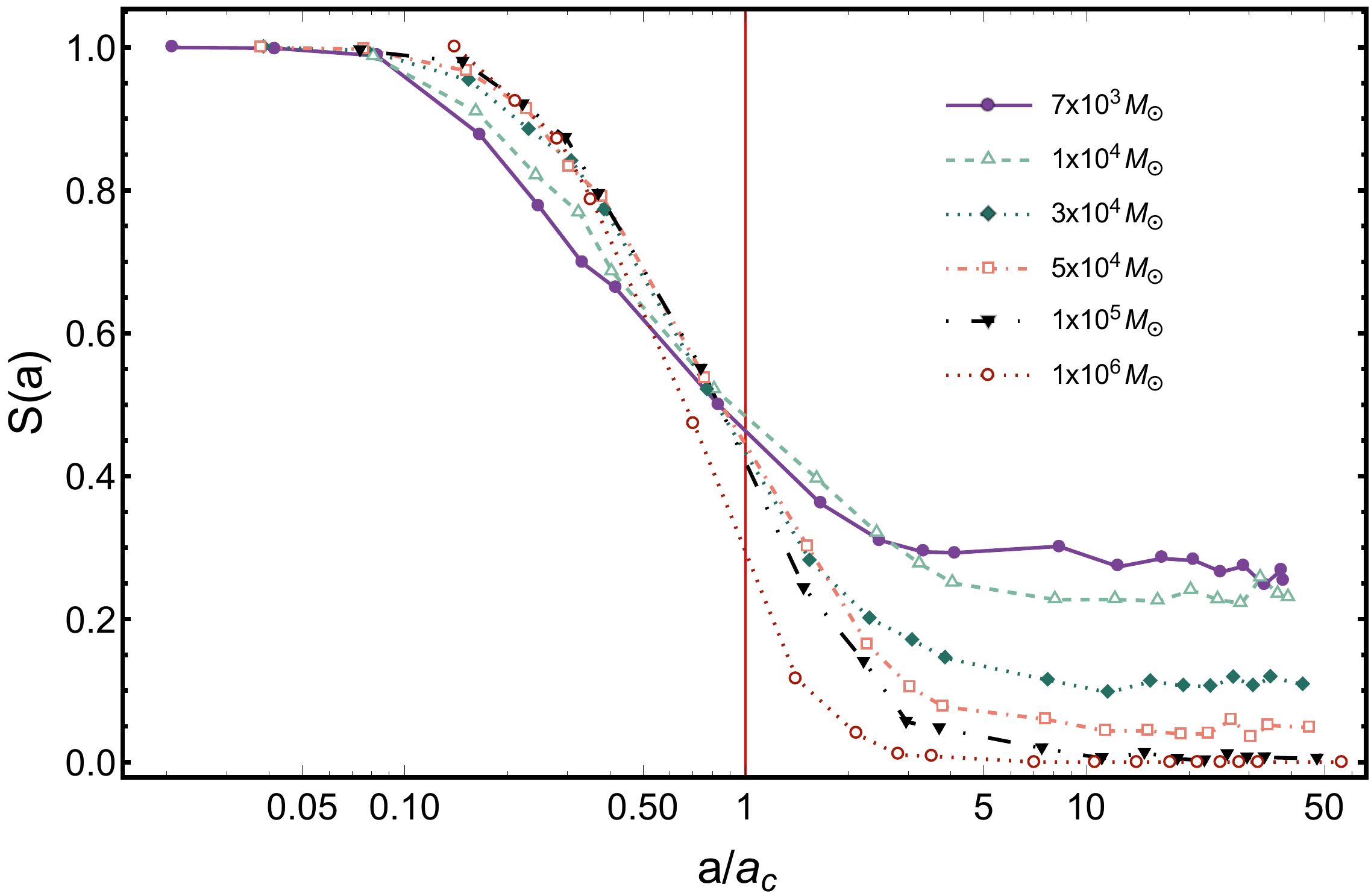}
\caption{The EMRI fraction $S(a)$ plotted against a dimensionless semimajor axis $a/a_{\rm c}$, for the inspiral of $m=10 M_\odot$ black holes through a stellar cusp ($\gamma = 7/4$) into a central MBH. Different curves correspond to different MBH masses $M_\bullet$ as labeled in the caption.  For $M_\bullet \gtrsim 10^5 M_\odot$, we recover the classic EMRI/plunge dichotomy, but cliffhanger EMRIs emerge for $M_\bullet \lesssim 10^5 M_\odot$ IMBHs (i.e. the EMRI fraction $S \sim \mathcal{O}(1)$ even as $a\to \infty$).}
\label{fig:SMass}
\end{figure}

\begin{figure}
\includegraphics[width=80mm]{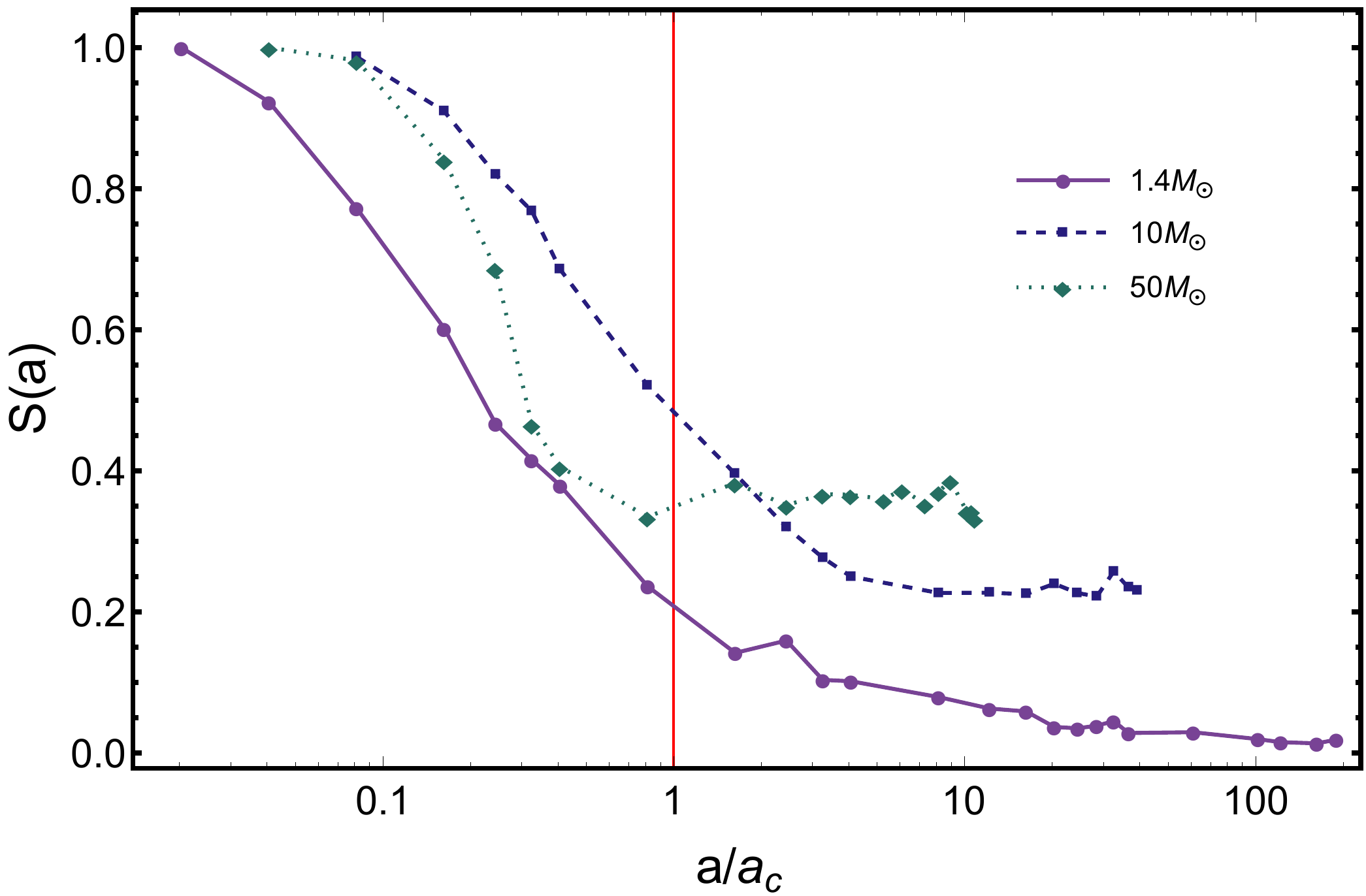}
\caption{Same as Fig. \ref{fig:SMass}, but now $M_\bullet = 10^4 M_\odot$ for all curves and the mass of the CO is varied (as labeled in the caption).  For $a > a_{\rm c}$, the cliffhanger EMRI fraction is higher for larger COs, and smaller for smaller COs.}
\label{fig:SCO}
\end{figure}

\begin{figure}
\includegraphics[width=80mm]{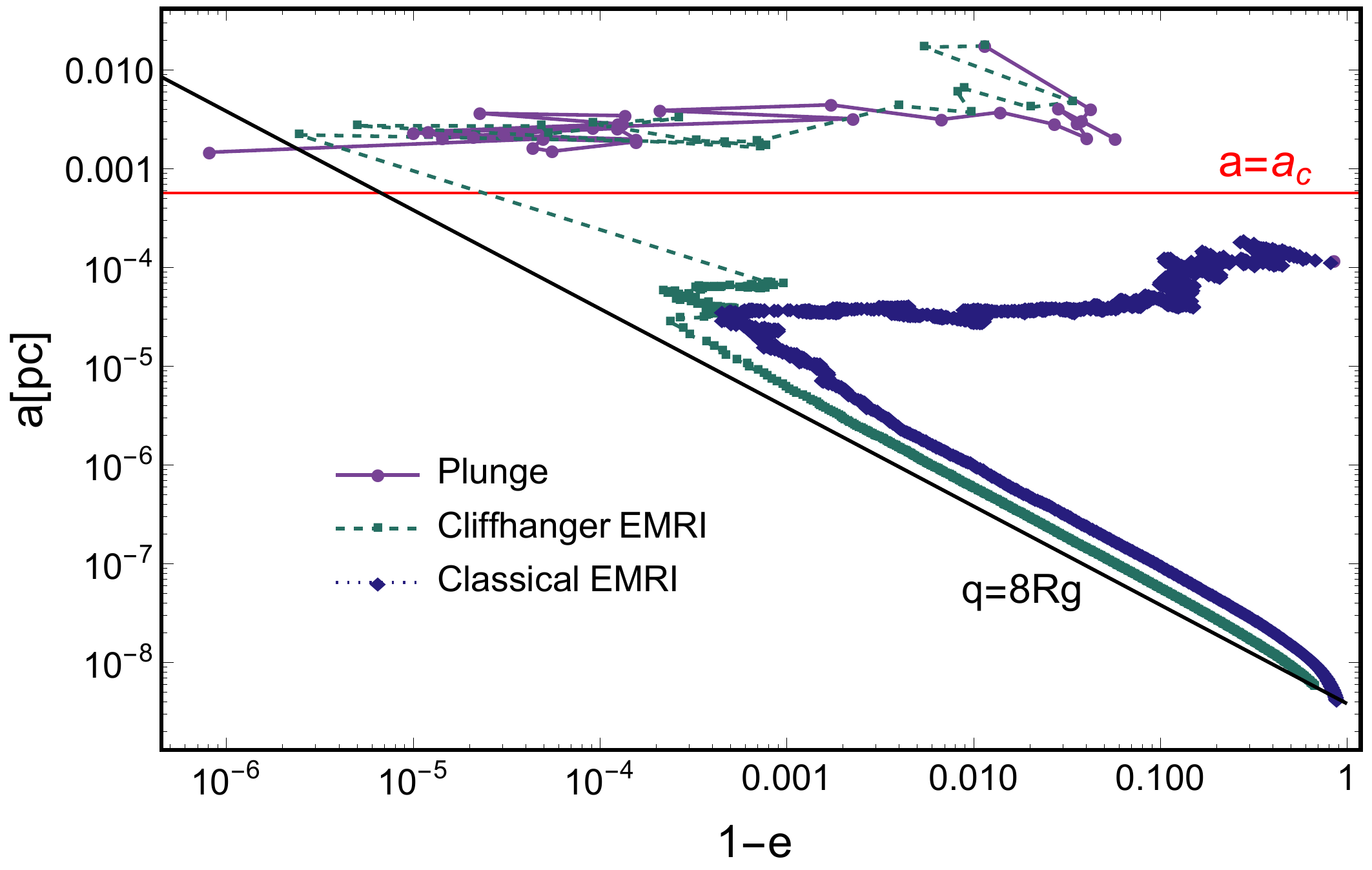}
\caption{Representative random walks in the space of angular momentum and energy, here represented as $\{1-e, a\}$.  The figure illustrates a plunge, a classical EMRI (with initial semimajor axis $a_0 < a_{\rm c}$), and a novel, cliffhanger EMRI (with $a_0 \gg a_{\rm c}$). }
\label{fig:random}
\end{figure}

To illustrate these alternate regimes more explicitly, we plot the random walks of different COs in Fig. \ref{fig:random} as seen in our MC calculations.  This diagram shows the MC trajectory of three $m=10M_\odot$ particles through the space of energy and angular momentum (parametrized as $a$ and $1-e$, respectively).  In addition to a standard plunge ($a_0 \gg a_{\rm c}$) and a classical EMRI ($a_0 \ll a_{\rm c}$), we see the explicit trajectory of a novel cliffhanger EMRI, which begins with $a_0 \gg a_{\rm c}$ but becomes an EMRI after a very close failed plunge dissipates so much energy that $a$ moves below $a_{\rm c}$.  This trajectory shows that cliffhanger EMRIs are fundamentally produced by non-diffusive orbital evolution, which (i) recalls other recent examples of non-diffusive effects in loss cone physics \citep{WeissbeinSari17, Teboul+22}, and (ii) may make them challenging to capture in a Fokker-Planck framework.

\begin{figure}
\includegraphics[width=80mm]{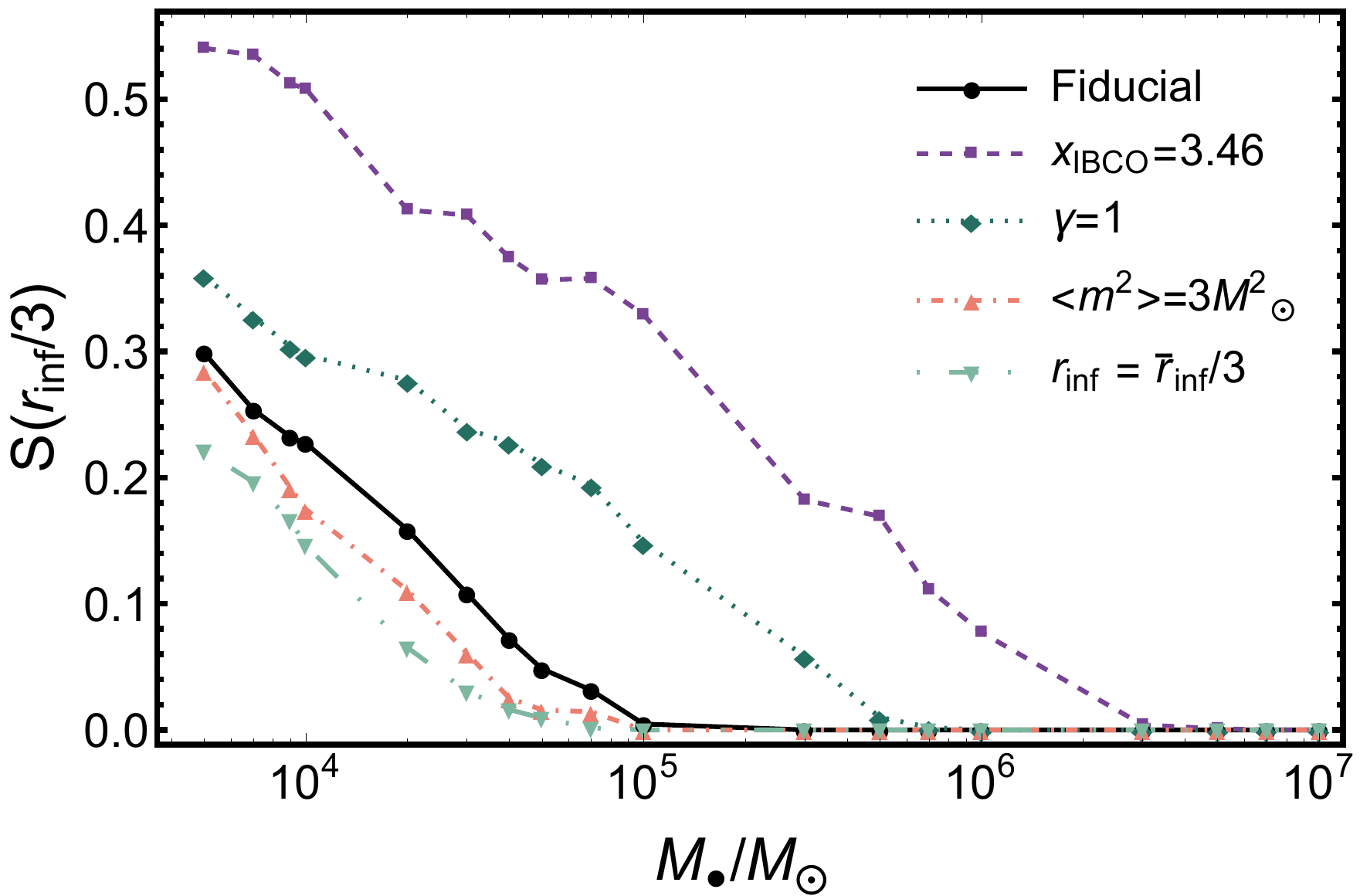}
\caption{The asymptotic EMRI fraction $S$ at values of $a \gg a_{\rm c}$ for an $m=10 M_\odot$ CO (here we take $a = r_{\rm inf}/3$ for consistency with our MC assumptions).  As predicted, $S(r_{\rm inf}/3) \to 0$ for $M \gtrsim M_{\rm c}$, which equals $10^{4.9}M_\odot$ for our fiducial model ($\gamma=7/4$, $\langle m^2 \rangle = M_\odot^2$, $r_{\rm inf}=\bar{r}_{\rm inf}$, $x_{\rm IBCO}=8$).  We also illustrate four non-fiducial cases where one of these four parameters is varied.  Increasing $\langle m^2 \rangle$ or decreasing the influence radius will decrease $M_{\rm c}$, while flattening the cusp (reducing $\gamma$) will increase it.  We crudely estimate the impact of rapid MBH spin by considering an $x_{\rm IBCO}=3.46$ case (appropriate for the equatorial plane geodesics of $\chi_\bullet = 0.9$ MBHs), which massively increases $M_{\rm c}$.}
\label{fig:SInfinity}
\end{figure}

In Fig. \ref{fig:SInfinity}, we test our analytic understanding of cliffhanger EMRIs by measuring a proxy for $S_\infty$ across a range of different parameter values (since our MC simulations assume a Keplerian potential, we take $S(r_{\rm inf}/3)$ as a reasonable approximation of $S_\infty$).  In this figure, we see first that Eq. \ref{eq:Mc} provides an excellent estimate for the maximum $M_\bullet$ value for which cliffhanger EMRIs occur. The value of this cutoff $M_{\rm c}$ depends significantly on $\gamma$; for $m=10 M_\odot$, $M_{\rm c}=10^{4.9}M_\odot$ ($M_{\rm c}=10^{5.5}M_\odot$) in a $\gamma=7/4$ ($\gamma=1$) cusp.  It depends more weakly on the value of $r_{\rm inf}$ and $\langle m_\star^2 \rangle$.  However, $M_{\rm c}$ can increase by 1-2 orders of magnitude if $x_{\rm IBCO}$ is reduced to values appropriate for prograde equatorial geodesics around rapidly spinning MBHs, which we discuss more below.

\noindent {\it Discussion}-- There are many uncertainties that currently prevent accurate computation of EMRI rates in our Universe. Most notably, there is not yet agreement on which of the various channels dominates total EMRI production. Even if we restrict ourselves to the classical loss cone channel for making EMRIs, the volumetric rate suffers from about three orders of magnitude of uncertainty \cite{Babak+17}, stemming primarily from (i) the unknown mass distribution of IMBHs in galactic nuclei, which likely dominate EMRI rates if they exist; (ii) the unknown density profile of stellar mass COs in galactic nuclei; and (iii) the rate at which COs are repopulated over cosmic time.  All three uncertainties are greatest in dwarf galactic nuclei, where very little is known about the IMBH mass function or stellar density profiles.  This is unfortunate, because EMRI rates (in the loss cone channel) are likely dominated by the smallest dwarf galaxies with a high occupation fraction.

EMRI rates from an individual galaxy can be approximated \citep{HopmanAlexander05} as
\begin{equation}
    \dot{N}_{\rm EMRI} = \int_0^{a_{\rm max}} \frac{N(a)S(a){\rm d}a}{\ln(J_{\rm c}/J_{\rm IBCO}) t_{\rm r}(a)},
\end{equation}
where $N(a)=4\pi a^2 n(a)$ and the energy relaxation time $t_{\rm r} = t_{\rm AM}/(1-e^2)$.  In past work \citep{HopmanAlexander05, RavehPerets21}, EMRI rates are generally calculated by setting $S(a)=1$ and $a_{\rm max} = a_{\rm c}$, i.e. assuming a step-function transition between EMRIs and plunges at $a=a_{\rm c}$.  Likewise, plunge rates can been roughly calculated as
\begin{equation}
    \dot{N}_{\rm pl} = \int_{a_{\rm min}}^\infty \frac{N(a)(1-S(a)){\rm d}a}{\ln(J_{\rm c}/J_{\rm IBCO}) t_{\rm r}(a)},
\end{equation}
with the past approximation that $S(a)=0$, $a_{\rm min}=a_{\rm c}$, and (sometimes) that the upper limit of integration can be replaced with $r_{\rm inf}$, since loss cone flux generally declines swiftly outside the influence radius \citep{StoneMetzger16}.
Since we have found that these approximations do not hold for small $M_\bullet$ or large $m$, we now revisit the issue of volumetric EMRI rates $\dot{n}_{\rm EMRI}$.  Our ultimate goal is to determine the plunge-to-EMRI ratio $\mathcal{R}=\dot{n}_{\rm pl}/\dot{n}_{\rm EMRI}$ (where $\dot{n}_{\rm pl}$ is the volumetric plunge rate).  

We estimate the total volumetric EMRI rate $\dot{n}_{\rm EMRI}$ in the following simplified way: 
\begin{equation}
    \dot{n}_{\rm EMRI} = \int_{\rm M_{\rm min}}^{\rm M_{\rm max}}\phi_\bullet(M_\bullet) \dot{N}_{\rm EMRI} (M_\bullet) {\rm d}M_\bullet. \label{eq:nDotEMRI}
\end{equation}
Here, $\phi_\bullet = {\rm d}n/{\rm d} M_\bullet$ is the differential volume density of massive black holes, which depends on $f_{\rm occ}(M_\bullet) \le 1$, the uncertain IMBH occupation fraction in dwarf galaxies \cite{StoneMetzger16}.  The minimum and maximum MBH masses relevant for {\it LISA}-band EMRIs are approximated as $M_{\rm min}\approx 10^4 M_\odot$ and $M_{\rm max} \approx 10^{6.5}M_\odot$, respectively.  A more complete description of how we estimate $\phi_\bullet(M_\bullet)$ and $f_{\rm occ}(M_\bullet)$ is included in the Supplementary Material, but in brief, we assume a steep decline in the occupation fraction for $M_\bullet < M_{\rm tr}$, an unknown transition mass.  We further assume $\gamma=7/4$, $\langle m_\star \rangle = 1 M_\odot$, and $\langle m_\star^2 \rangle = 1 M_\odot^2$.

\begin{figure}
\includegraphics[width=80mm]{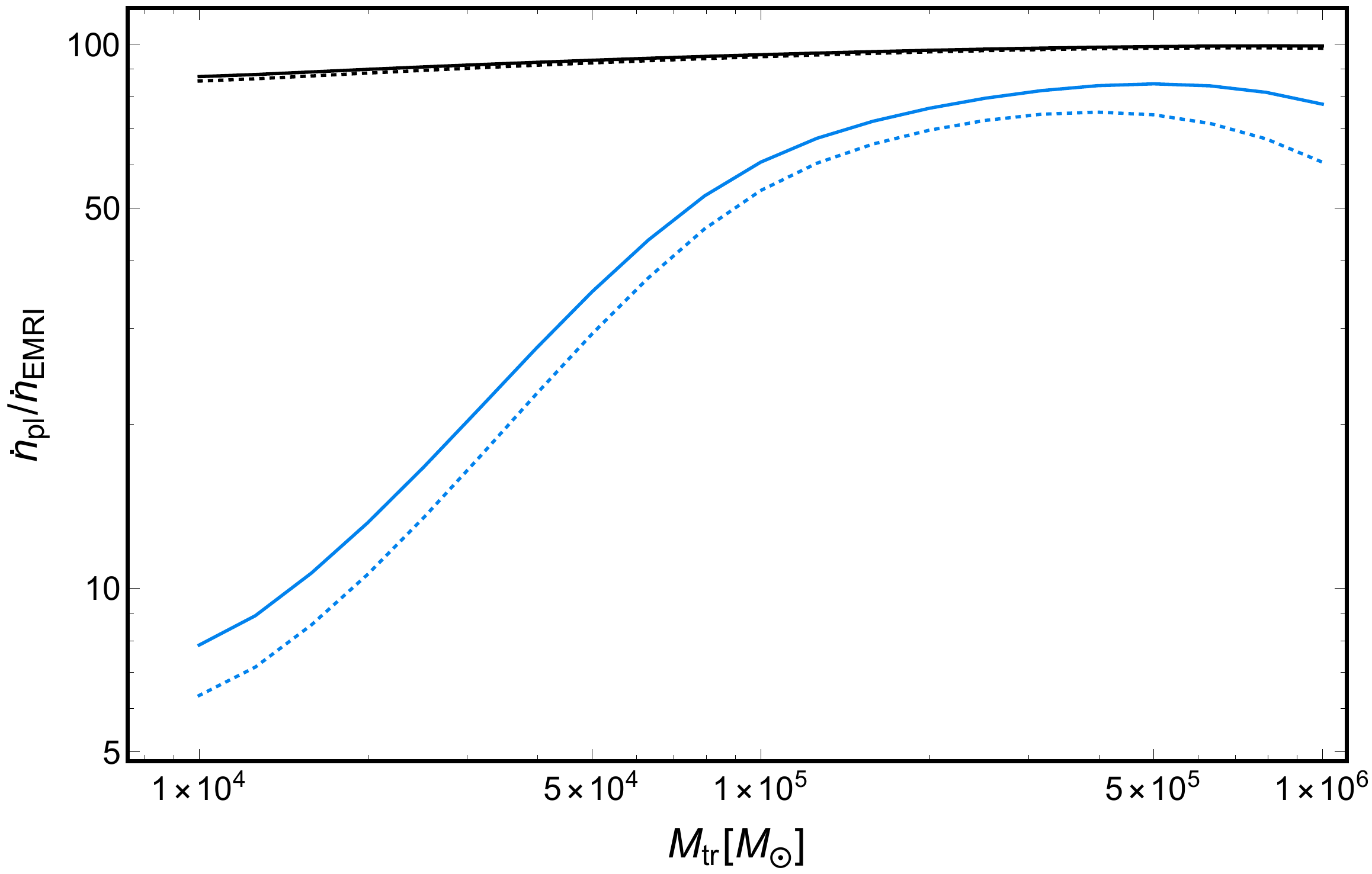}
\caption{The ratio of volumetric plunge to EMRI rates, $\mathcal{R}=\dot{n}_{\rm pl}/\dot{n}_{\rm EMRI}$, plotted against the unknown transition mass $M_{\rm tr}$ where the occupation fraction $f_{\rm occ}$ of MBHs in dwarf galaxies falls below $50\%$.  Black curves show $\mathcal{R}$ disregarding the existence of cliffhanger EMRIs, while blue curves show $\mathcal{R}$ including cliffhangers.  Solid lines show one estimate of the MBH mass function \cite{StoneMetzger16}, and dotted show another \cite{Shankar+09, Kochanek16} (see Supplementary Material). Notably, $R$ is largely independent of the slope of the MBH mass function, though it depends sensitively on its lower truncation $M_{\rm tr}$. If Schwarzschild IMBHs are abundant in the Universe, cliffhanger EMRIs increase the volumetric EMRI rate by roughly one order of magnitude.}
\label{fig:rates}
\end{figure}

In Fig. \ref{fig:rates}, we show that $\mathcal{R}$ depends sensitively on $M_{\rm tr}$.  If $M_{\rm tr}\gtrsim 10^5 M_\odot$, then $\mathcal{R} \sim 100$, in agreement with past estimates \cite{BarOrAlexander16}.  However, if  $M_{\rm tr}\sim 10^4 M_\odot$, then the plunge-to-EMRI ratio can drop to $\mathcal{R}\sim 10$ or even lower, representing a 1 order of magnitude increase in volumetric EMRI rates from the (previously neglected) contribution of cliffhanger EMRIs.  This conclusion is likely conservative, however, because our MC simulations have focused on Schwarzschild MBHs.  The addition of spin to the problem can significantly change EMRI dynamics.  For prograde orbits around a rapidly spinning black hole, $\mathcal{R}$ can decrease dramatically, due to the drop in $x_{\rm IBCO}$ and the concomitantly greater ``window of opportunity'' for GW dissipation to counteract two-body relaxation \citep{AmaroSeoane+13, VazquezAceves+22}.  While this spin dependence has been studied in the past for standard EMRIs, it also clearly increases the mass range over which cliffhanger EMRIs can occur; decreasing $x_{\rm IBCO}$ will increase $M_{\rm c}$ (see Eq. \ref{eq:Mc}).  For example, if $m=10 M_\odot$, going from a Schwarzschild black hole ($x_{\rm IBCO}=8$) to the extremal Kerr prograde case ($x_{\rm IBCO}=2$) increases $M_{\rm c}$ from $M_{\rm c}\approx 10^{4.9}M_\odot$ up to $M_{\rm c}\approx 10^{7.1}M_\odot$.  We have shown a test calculation of the effects of spin in Fig. \ref{fig:SInfinity}, but caution that this is a very crude estimate, as it neglects both inclination diffusion and higher-order post-Newtonian corrections to GW losses.  Indeed, as MBH spin increases and $x_{\rm IBCO}$ becomes more relativistic, a post-Newtonian treatment may become wholly inappropriate, necessitating fully relativistic calculations.  A full exploration of spin effects is beyond the scope of this Letter, but may plausibly push the importance of cliffhanger EMRIs from the IMBH into the SMBH regime.

Here we have only calculated volumetric rates ($\dot{n}_{\rm EMRI}$, $\dot{n}_{\rm pl}$) and rates ratios $\mathcal{R}$; a more complete exploration of {\it LISA} detection rates is also beyond the scope of this paper.  However, as {\it LISA} retains significant sensitivity to EMRIs for $10^4 M_\odot \lesssim M_\bullet \lesssim 10^5 M_\odot$, it is plausible that a large change in $\mathcal{R}$ produced by cliffhanger EMRIs could increase {\it LISA} detection rates substantially. {\it LISA} detection rates would certainly increase if most MBHs are rapidly spinning and the cliffhanger dynamics presented here extend up to $M_\bullet \sim 10^6 M_\odot$.

To the best of our knowledge, cliffhanger EMRIs have not been seen in past MC simulations.  In some cases this is because sufficiently small IMBHs were not simulated \citep{HilsBender95, RavehPerets21}, but in other cases, this novel EMRI population was not seen even for $M=10^{3-4} M_\odot$ \citep{HopmanAlexander05, BarOrAlexander16}.  In the case of Ref. \citep{HopmanAlexander05}, this is likely due to a sign error on the MC implementation of $\langle \Delta L \rangle_{\rm o}$ \citep{RavehPerets21}.  For Ref. \citep{BarOrAlexander16}, this seems to be due to the choice to simulate cusps where every particle has $m=10M_\odot$, which pushes $M_{\rm c}$ down to roughly $10^4 M_\odot$. 

Finally, we note that while the main focus of this work has been on individual EMRIs, a cosmological population of unresolved EMRIs may also be an important source of signal (or noise) for {\it LISA}.  The magnitude of this stochastic EMRI background is also dominated by true EMRIs around the smallest MBHs with a large volume density \citep{Pozzoli+23}, and may in practice be dominated by the cliffhanger EMRIs identified in this paper.

\begin{acknowledgments}
We gratefully acknowledge helpful conversations with Luca Broggi, Jenny Greene, Barak Rom, Re'em Sari and Marta Volonteri.  We are especially grateful to Cole Miller, whose seminar on EMRIs helped us to understand and to formulate an analytic interpretation of $M_{\rm c}$ and $x_{\rm c}$, and who also provided constructive comments on an earlier version of this paper.  IQ and NCS acknowledge financial support from  the Israel Science Foundation (Individual Research Grant 2565/19).  NCS is further supported by the Binational Science foundation (grant Nos. 2019772 and 2020397).
\end{acknowledgments}

\bibliography{ms}

\appendix

\section{Diffusion Coefficients}
\label{app:diff}
We compute local and orbit-averaged diffusion coefficients in the standard way, see. e.g. Ref. \cite{Merritt13} for a pedagogical summary.  For completeness, we enumerate our definitions here.  The orbit-averaged diffusion coefficients used in the MC calculations are

\begin{equation}
\langle \Delta E \rangle_{\rm o} = 2m\int_{r_{p}}^{r_{a}} \langle \Delta \epsilon \rangle {\rm d}r/v_{\rm r} ,
\end{equation}

\begin{equation}
\langle (\Delta E)^2 \rangle_{\rm o} =  2m\int_{r_{p}}^{r_{a}} \langle (\Delta \epsilon)^2 \rangle {\rm d}r/v_{\rm r} ,
\end{equation}

\begin{equation}
\langle \Delta L \rangle_{\rm o} = 2m\int_{r_{p}}^{r_{a}} \langle \Delta J \rangle  {\rm d}r/v_{r} ,
\end{equation}

\begin{equation}
\langle (\Delta L)^2 \rangle_{\rm o} =  2m\int_{r_{p}}^{r_{a}} \langle (\Delta J)^2 \rangle {\rm d}r/v_{\rm r},
\end{equation}

\begin{equation}
\langle \Delta E \Delta L \rangle_{\rm o} =  2m\int_{r_{p}}^{r_{a}}\langle \Delta \epsilon \Delta J \rangle {\rm d}r/v_{\rm r},
\end{equation}

where $v_{\rm r}$ is the radial velocity, $r_{\rm p}$ is the orbital pericenter, and $r_{\rm a}$ is the orbital apocenter.  $\langle \Delta \epsilon \rangle$, $\langle (\Delta \epsilon)^2 \rangle$, $\langle \Delta J \rangle$, $\langle (\Delta J)^2 \rangle$, and $\langle \Delta \epsilon \Delta J \rangle$ are local specific energy and specific angular momentum diffusion coefficients related to the local velocity diffusion coefficients in the following way:
\begin{equation}
\langle \Delta \epsilon \rangle = v \langle \Delta v_{\parallel} \rangle + \frac{1}{2} \langle (\Delta v_{\parallel})^2 \rangle + \frac{1}{2} \langle (\Delta v_{\perp})^2 \rangle
\end{equation}

\begin{equation}
\langle (\Delta \epsilon)^2 \rangle = v^2 \langle (\Delta v_{\parallel})^2 \rangle
\end{equation}

\begin{equation}
J \langle \Delta J \rangle = \frac{J^2}{v} \langle \Delta v_{\parallel} \rangle + \frac{r^2}{4} \langle (\Delta v_{\perp})^2 \rangle
\end{equation}

\begin{equation}
\langle (\Delta J)^2 \rangle = \frac{J^2}{v^2} \langle (\Delta v_{\parallel})^2 \rangle + \frac{1}{2}\left(r^2-\frac{J^2}{v^2}\right) \langle (\Delta v_{\perp})^2 \rangle
\end{equation}

\begin{equation}
\langle \Delta \epsilon \Delta J \rangle = J \langle (\Delta v_{\parallel})^2 \rangle.
\end{equation}

Here $\langle \Delta v_{\parallel} \rangle$, $\langle (\Delta v_{\parallel})^2 \rangle$, and $\langle (\Delta v_{\perp})^2 \rangle$ are the local velocity diffusion coefficients, defined as
\begin{equation}
\langle \Delta v_{\parallel} \rangle = -16 \pi^2 G^2 (\langle m^2 \rangle + m\langle m \rangle)  \ln\Lambda F_{2}(v)
\end{equation}

\begin{equation}
\langle (\Delta v_{\parallel})^2 \rangle = \frac{32 \pi^2}{3} G^2 \langle m^2 \rangle \ln\Lambda v [F_4(v) + E_1(v)]
\end{equation}

\begin{equation}
\langle (\Delta v_{\perp})^2 \rangle = \frac{32 \pi^2}{3} G^2 \langle m^2 \rangle \ln\Lambda v [3F_2(v) -F_4(v) + 2E_1(v)]
\end{equation}

with the functions $E_{n}(v)$ and $F_{n}(v)$ given by
\begin{equation}
E_{n}(v) = \int_{0}^{\infty}(\frac{v_f}{v})^{n} f(v_f) {\rm d}v_f
\end{equation}

\begin{equation}
F_{n}(v) = \int_{0}^{v}(\frac{v_f}{v})^{n} f(v_f) {\rm d}v_f.
\end{equation}
We numerically implement orbital diffusion as described in the main {\it Letter}, dynamically choosing a timestep that covers a number of orbits $N\ge 1$.  We set $N=\max(1, N_{\rm safe})$ where $N_{\rm safe} = (1/F_{\rm safe}) \min(E / \Delta E_{\rm GW}, \sqrt{L^2 / (\langle (\Delta L)^2 \rangle_{\rm o}P)})$, where $P$ is the orbital period.  We use $F_{\rm safe}=10$ by default, but we have tested the convergence of our results with alternate $F_{\rm safe}=3$ and $F_{\rm safe}=30$.  Our final $S(a)$ curves are not sensitive to choice of $F_{\rm safe}$.

\begin{figure}
\includegraphics[width=80mm]{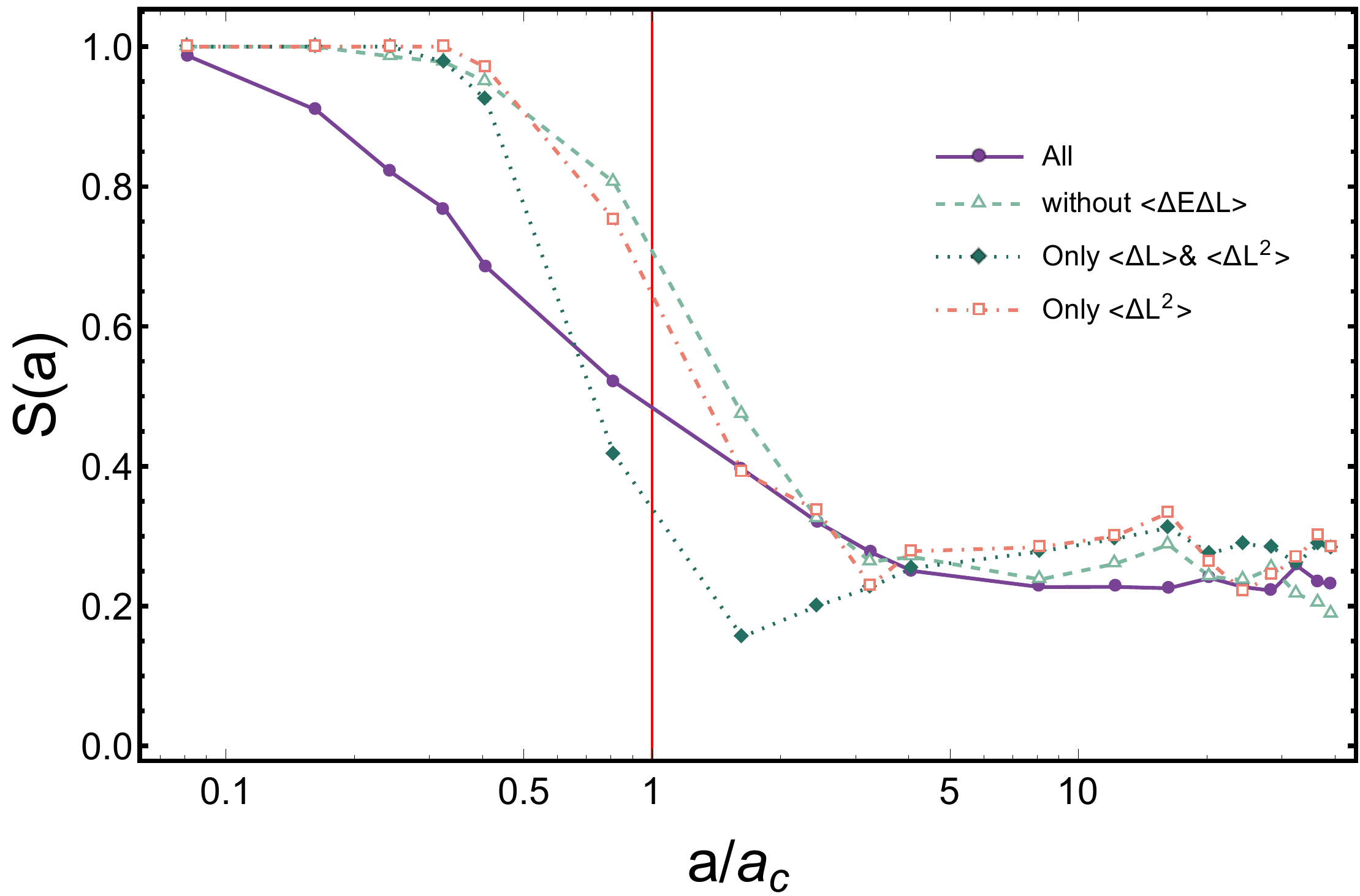}
\caption{The EMRI fraction $S(a)$ plotted against a normalized semimajor axis $a$.  Here we show $S$-curves for $M=10^4 M_\odot$, $m = 10 M_\odot$, and $\gamma=7/4$, but experiment by turning off different combinations of diffusion coefficients.  We see that the accurate calculation transitions from $S\approx 1$ to $S\approx S_\infty$ at a smaller semimajor axis than in any of the numerical experiments where we turn different combinations of diffusion coefficients off.  In no cases do these experiments eliminate the population of cliffhanger EMRIs, demonstrating that this novel EMRI source comes from the interplay of $\Delta E_{\rm GW}$ and $\langle (\Delta L)^2 \rangle_{\rm o}$ around IMBHs, as we argued in the main text.}
\label{fig:DCs}
\end{figure}

Past MC calculations of EMRI properties have generally used only a limited set of the above diffusion coefficients; for example, Refs. \cite{HopmanAlexander05, RavehPerets21} only employ GW dissipation, $\langle \Delta L \rangle_{\rm o}$, and $\langle (\Delta L)^2 \rangle_{\rm o}$.  In contrast, Ref. \cite{BarOrAlexander16} used the full set of diffusion coefficients enumerated here, as well as additional ones encoding the effect of scalar resonant relaxation (though this source of relaxation was found to be ultimately unimportant for EMRI dynamics).  We examine the importance of different diffusion coefficients in Fig. \ref{fig:DCs}.  While selectively turning off individual diffusion coefficients changes the quantitative shape of the $S(a)$ curve, particularly near $a_{\rm c}$, so long as $\langle (\Delta L)^2 \rangle_{\rm o}$ is allowed to compete with GW dissipation, the qualitative and asymptotic behavior is unchanged.  In particular, the cliffhanger EMRI phenomenon appears to be robust to the inclusion or exclusion of diffusion coefficients other than $\langle (\Delta L)^2 \rangle_{\rm o}$.

\section{Volumetric Rates}
\label{app:volume}
The volumetric event rates we compute ($\dot{n}_{\rm EMRI}$ and $\dot{n}_{\rm pl}$) both depend sensitively on the MBH mass function $\phi_\bullet(M_\bullet)$, and indeed are dominated by the smallest galaxies with a high MBH occupation fraction (at least under the assumptions employed in this work).  Because of the broad uncertainties related to the bottom end of the MBH mass function, we consider two different approaches to estimating it, each of which is motivated by analogous calculations of tidal disruption event rates \citep{StoneMetzger16, Kochanek16}.

In the first approach \cite{StoneMetzger16}, we begin with the usual galaxy luminosity function 
\begin{equation}
    \phi(L) = \frac{n_0}{L_0} \left( \frac{L}{L_0} \right)^\alpha \exp(-L/L_0).
\end{equation}
More specifically, we employ a calibration of the K-band galaxy luminosity function with $\alpha=-1.09$, $n_0 = 1.16\times 10^{-2} h^3~{\rm Mpc}^{-3}$, and $L_0 = 5.15\times 10^{10}L_\odot$ \cite{Kochanek+01}.
We proceed by setting $\tilde{\phi}_\bullet(M_\bullet) = \phi(L) ({\rm d}L/{\rm d}M_\bullet$), where $L(M_\bullet)$ is estimated by combining the Faber-Jackson law,
\begin{equation}
L = 10^{10} L_\odot \left( \frac{\sigma_{\rm e}}{150~{\rm km~s}^{-1}} \right)^{4}
\end{equation}
with a recent calibration of the $M_\bullet-\sigma$ relationship \citep{Tremaine+02} employing upper limits on $M_\bullet$ in dwarf galaxies \cite{Greene+20},
\begin{equation}
    \log_{10}(M_\bullet/M_\odot) = 7.87+4.55\log_{10}\left( \frac{\sigma_{\rm e}}{160~{\rm km~s}^{-1}} \right).
\end{equation}
Finally, we convert $\tilde{\phi}_\bullet$, which represents an upper limit on the MBH mass function assuming $100\%$ occupancy of all dwarf nuclei, into the true MBH mass function via $\phi_\bullet(M_\bullet) = \tilde{\phi}(M_\bullet) f_{\rm occ}(M_\bullet)$.  We use an ad hoc occupation fraction parametrization \cite{Miller+15}
\begin{equation}
    f_{\rm occ} = \frac{1}{2}+\frac{1}{2}\tanh\left(\ln\left(\frac{M_\bullet}{M_{\rm tr}}\times 2.5^{5.9-\log_{10}(M_{\rm tr}/M_\odot)} \right) \right)
\end{equation}
that satisfies (i) $f_{\rm occ} \to 1$ as $M_\bullet/M_{\rm tr} \to \infty$, (ii) $f_{\rm occ} \to 0$ as $M_\bullet/M_{\rm tr} \to 0$, and (iii) $f_{\rm occ}=1/2$ when $M_\bullet = M_{\rm tr}$, a transition mass which is the free parameter in this function.
With $\phi_\bullet(M_\bullet)$ thus defined, we use Eq. \ref{eq:nDotEMRI} to compute $\dot{n}_{\rm EMRI}$, interpolating our MC results over the space of $\{a, M_\bullet\}$ to compute $S(a, M_\bullet)$.  We compute the volumetric plunge rate $\dot{n}_{\rm pl}$ in the same way (replacing $\dot{N}_{\rm EMRI}$ with $\dot{N}_{\rm pl}$), and finally determine the total plunge-to-EMRI ratio $\mathcal{R}=\dot{n}_{\rm pl}/\dot{n}_{\rm EMRI}$. 

In the second approach \cite{Kochanek16}, we take empirically calibrated models for MBH populations that link MBH growth rates with observed active galactic nucleus luminosity functions \citep{Shankar+09}.  Again, we assume that these tabulated models (which extend to a minimum $M_\bullet = 1\times 10^5 M_\odot$, and which we extrapolate as power laws to lower masses) are really only models for $\tilde{\phi}_\bullet(M_\bullet)$, and we correct this into the true MBH mass function as above ($\phi_\bullet = \tilde{\phi}_\bullet f_{\rm occ}$).  

Ultimately, both of these methods give uncertain and debatable approximations for the unknown MBH mass function in dwarf galaxies.  However, our results (Fig. \ref{fig:rates}) show that the shape of $\tilde{\phi}_\bullet$ does not have a leading-order impact on $\mathcal{R}$, which is affected much more by $f_{\rm occ}$ (or, in our parametrization, $M_{\rm tr}$).  Better models for $\tilde{\phi}_\bullet$ would be needed to accurately assess the true volumetric EMRI rate, $\dot{n}_{\rm EMRI}$, however.

\end{document}